\documentclass[conference]{IEEEtran}
\IEEEoverridecommandlockouts
\usepackage{cite}
\usepackage{amsmath,amssymb,amsfonts}
\usepackage{algorithmic}
\usepackage{graphicx}
\usepackage{textcomp}
\usepackage[table]{xcolor}
\definecolor{DarkGreen}{rgb}{0.0, 0.5, 0.0} 
\usepackage{gensymb}
\usepackage{array} 
\usepackage{soul}
\usepackage{multirow}
\begin{document}

\title{A nonlinear real time capable motion cueing algorithm based on deep reinforcement learning\\
}

\author{\IEEEauthorblockN{Hendrik Scheidel}
	\IEEEauthorblockA{\textit{Institute of System Dynamics and Control, Space Systems Dynamics,}\\
		\textit{Deutsches Zentrum für Luft- und Raumfahrt (DLR)}\\
		\textit{German Aerospace Center}\\
		82234, Germany\\
		hendrik.scheidel@dlr.de}
}

\author{
	Hendrik Scheidel,
	Camilo Gonzalez,
	Houshyar Asadi, \IEEEmembership{Member, IEEE},
	Tobias Bellmann, 
	Andreas Seefried, \\
	Shady Mohamed, 
	Saeid Nahavandi, \IEEEmembership{Fellow, IEEE}
	\thanks{This work has been submitted to the IEEE for possible publication. Copyright may be transferred without notice, after which this version may no longer be accessible. \linebreak}
	\thanks{H. Scheidel, C. Gonzalez, H. Asadi and S. Mohamed are with the Institute
		for Intelligent Systems Research and Innovation, Deakin University,
		Geelong, VIC 3216, Australia 
		(e-mail: hscheidel@deakin.edu.au; c.gonzalezarango@deakin.edu.au; houshyar.asadi@deakin.edu.au; shady.mohamed@deakin.edu.au).}
	\thanks{H. Scheidel, T. Bellmann, and A. Seefried are with the Institute of Flight Systems, German Aerospace Center, 
		82234 Weßling, Germany 
		(e-mail: hendrik.scheidel@dlr.de; tobias.bellmann@dlr.de; andreas.seefried@dlr.de).}
	\thanks{S. Nahavandi is with Swinburne University,
		Melbourne, VIC 3122, Australia 
		(e-mail: snahavandi@swin.edu.au).}
}

\maketitle

\begin{abstract}
In motion simulation, motion cueing algorithms are used for the trajectory planning of the motion simulator platform (MSP), where workspace limitations prevent direct reproduction of reference trajectories. Strategies such as motion washout, which return the platform to its center, are crucial in these settings. For serial robotic MSPs with highly nonlinear workspaces, it is essential to maximize the efficient utilization of the MSP’s kinematic and dynamic capabilities. Traditional approaches, including classical washout filtering and linear model predictive control, fail to consider platform-specific, nonlinear properties, while nonlinear model predictive control (NMPC), though comprehensive, imposes high computational demands that hinder real-time, pilot-in-the-loop application without further simplification.

To overcome these limitations, we introduce a novel approach using deep reinforcement learning (DRL) for motion cueing, demonstrated here for the first time in a 6-degree-of-freedom (DOF) setting with full consideration of the MSP's kinematic nonlinearities. 
Previous work by the authors successfully demonstrated the application of DRL to a simplified 2-DOF setup, which did not consider kinematic or dynamic constraints. This approach has been extended to all 6 DOF by incorporating a complete kinematic model of the MSP into the algorithm, a crucial step for enabling its application on a real motion simulator. 

The training of the DRL-MCA is based on Proximal Policy Optimization (PPO) in an actor-critic implementation combined with an automated hyperparameter optimization. After detailing the necessary training framework and the algorithm itself, we provide a comprehensive validation, demonstrating that the DRL MCA achieves competitive performance against established algorithms. Moreover, it generates feasible trajectories by respecting all system constraints and meets real-time requirements with low computational demand. Finally, we present the successful application of the DRL MCA to a real robotic motion simulator with a driver-in-the-loop, highlighting its practical effectiveness.
\end{abstract}


\section{Introduction}

Motion cueing algorithms (MCAs) are essential for calculating the movements of motion simulator platforms (MSP) to enhance the immersion of passengers in a simulated environment~\cite{2008Fischer}. Due to the inherent limitations of the workspace of MSPs, it is often not possible to replicate the full range of vehicle motions. Therefore the use of a MCA is needed to compute a trajectory that maximizes the sensory perception of motion for the passenger, within these constraints.

\begin{figure}
	\includegraphics[width=\columnwidth]{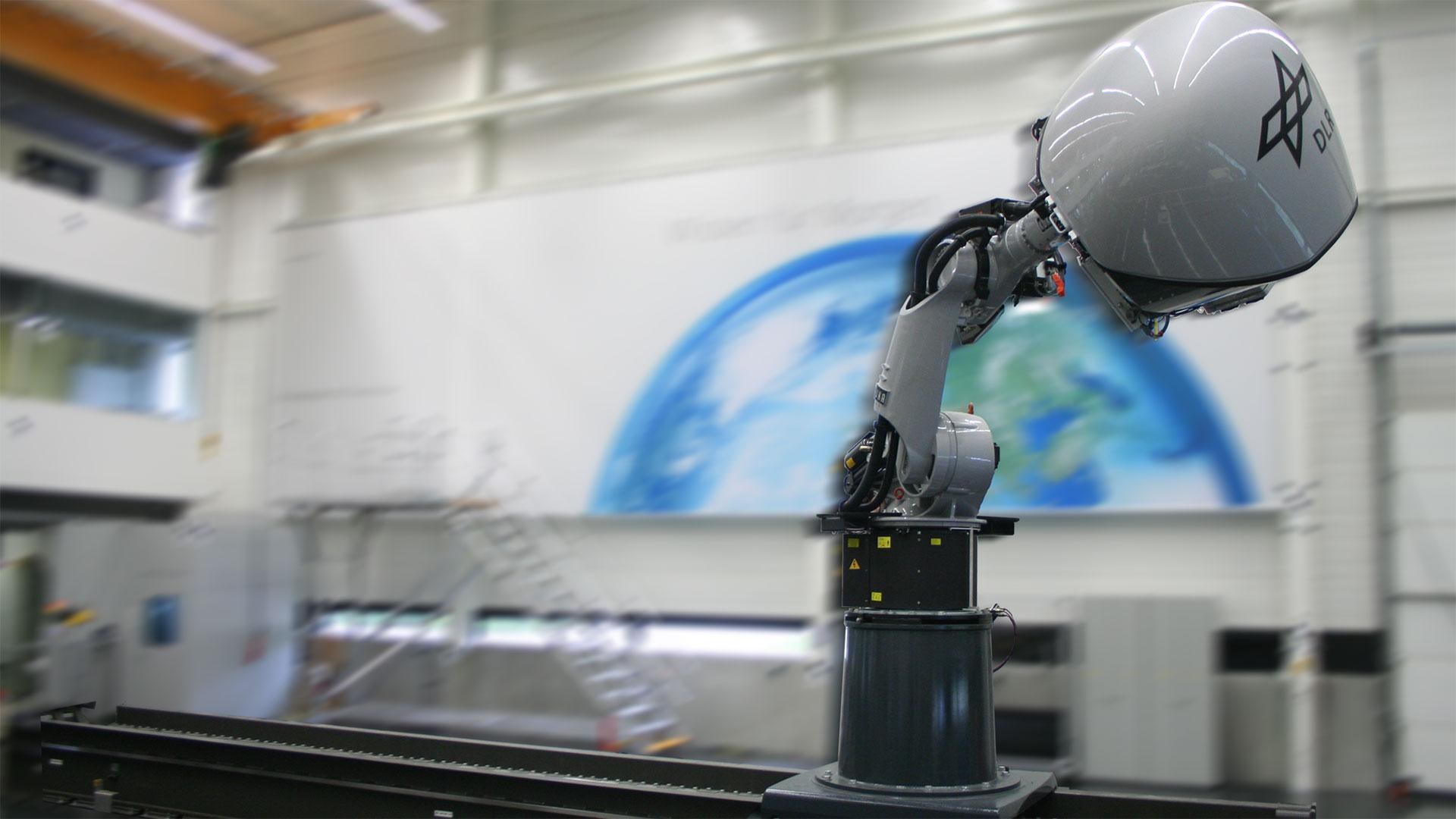}
	\caption{The DLR Robotic Motion Simulator~\cite{2011Bellmann_CONFa}.} 
	\label{fig:dlr_rms}
\end{figure}

This perception of motion is dominantly conducted by the human vestibular system. It consists of the otolith organs and semicircular canals, which are primarily perceiving the translational acceleration and angular velocity. Consequently, MCAs aim to reproduce these specific motion cues in order to maximize the passenger's immersion and to minimize the sensory discrepancy which can lead to motion sickness~\cite{2023Irmak_ARTC}.

Different types of motion platforms are employed in simulators, with the most common being parallel kinematic systems and serial kinematic systems based on industrial robotics. Parallel kinematic systems, often called hexapod, typically come with a higher payload but with a strongly limited workspace. Serial robotic platforms, as shown in Figure~\ref{fig:dlr_rms}, come with a increased workspace size which leads at the same time to the need of a more complex MCAs~\cite{2011Bellmann_CONF}. For parallel configurations, the workspace is often linearly approximated using simplified cubic shape. However, this approach is not applicable to serial robotic platforms, which possess a more complex, non-cubic workspace characterized by nonlinearities and singularities.  In robotic platforms, the positions, velocities, and accelerations of individual joints become critical limiting factors that must be considered to ensure accurate and realistic motion simulation.

\begin{figure}
	\includegraphics[width=\columnwidth]{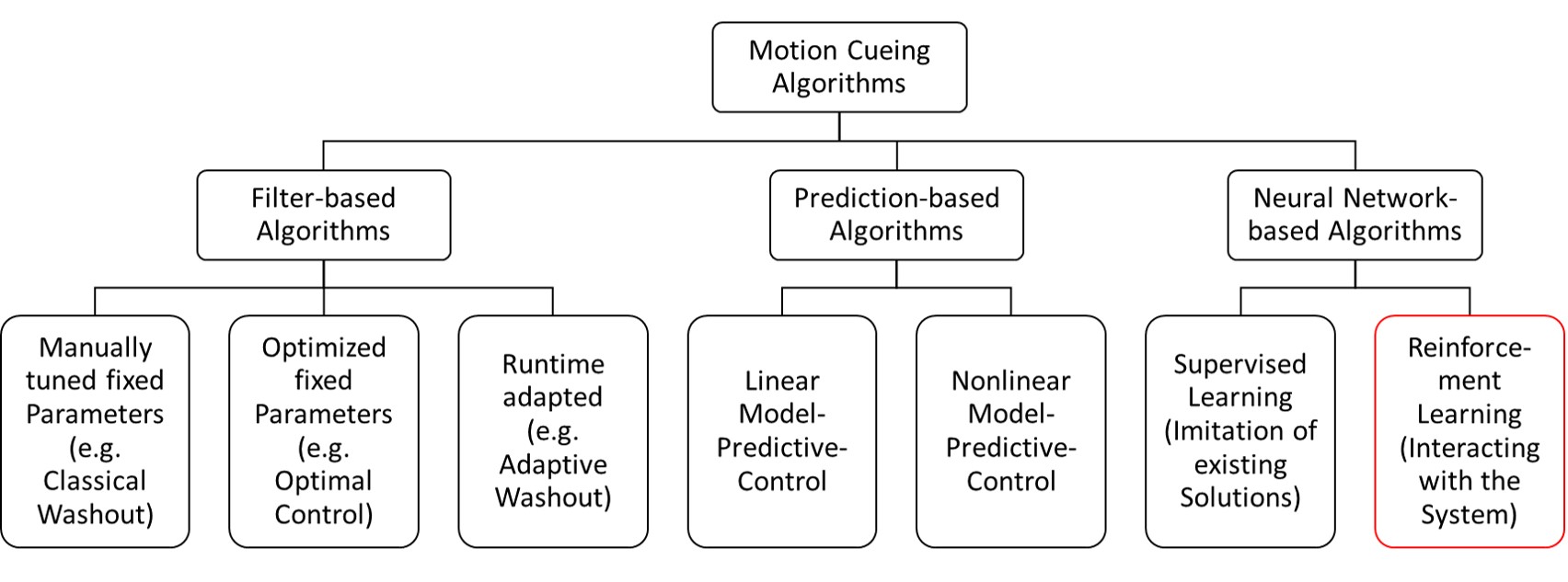}
	\caption{A proposed taxonomy of existing motion cueing algorithms. As this work is presenting a neural network-based algorithm achieved through reinforcement learning, this subcategory is marked in red.} 
	\label{fig:taxonomy}
\end{figure}

Various algorithms of different complexity have been developed for motion cueing. As shown in Figure~\ref{fig:taxonomy} they can be broadly classified into three categories: filter-based, model predictive, and neural network-based (NN) approaches. Each of these categories represents a different stage in the evolution of MCAs.

Filter-based algorithms are the earliest and most foundational, beginning with the classical washout (CW) algorithm~\cite{1985Reid_ARTC}, which filters translational and rotational accelerations to keep the simulator within workspace limits. Like most other algorithms, CW employs the concept of tilt coordination, in which slow pitch and roll tilting of the MSP's cabin is executed to align the gravitational force direction with the low-frequency or continuous components of the specific force from the reference. This is possible due to the somatogravic illusion~\cite{2006MacNeilage_ARTC}. The main advantage of filter-based approaches is their simplicity, which allows for fast execution. However, this simplicity also imposes significant limitations, often resulting in poor performance, due to fixed filter parameters and phase lag, or violations of kinematic constraints, particularly in complex nonlinear workspaces. 
More advanced filter based methods tried to attack this disadvantages, such as the optimal control algorithm, which minimizes errors between desired and actual sensory input based on a model of the human vestibular system, and the adaptive washout algorithm, which dynamically adjusts its parameters in real-time~\cite{2022Qazani_ARTC, 2022Asadi_ARTC}. These methods improved the capabilities of filter based motion cueing but the method is still limited to the relatively simple and rigid structure of filters.

The next major development is model predictive control (MPC), as introduced by Dagdelen et al.~\cite{2009Dagdelen_ARTC}. MPC is an optimal control scheme that is widely used in the process industry and robotics. Compared to other control schemes, its main advantages include seamless handling of MIMO systems, ease of consideration of state, input and output constraints, the guarantee of closed loop stability by design, and the ability to exploit knowledge and predictions of future control references or disturbances \cite{2021Schwenzer_ARTC, 2016Kouvaritakis}. In a nutshell, MPC controllers use a model or “plant" of the controlled system to find a sequence of control inputs that will result in optimal system response over a future period of time, or “horizon", given the current states of the system and desired reference trajectories, while explicitly handling constraints or limitations of the model. This is achieved by solving an optimization problem where the design variables are the sequence of future system inputs, and the plant is used to calculate future system outputs and states for said sequence. 
Linear MPC (LMPC)-based MCAs rely on a linear model of the MSP. When a hexapod is used as the MSP, approximating the workspace as a cube with linear characteristics can be a valid simplification. However, for more complex and nonlinear workspaces, such as those of a serial robotic simulator, this simplification can result in significant discrepancies between the model and reality, potentially leading to severe violations of the MSP's kinematic constraints. Nonlinear MPC (NMPC)-based MCAs, by contrast, account for the full nonlinear kinematics of a serial robot in their model. This increased accuracy comes with high computational costs, as solving a nonlinear program at every time step is required. As a result, real-time implementation is often either infeasible or requires additional simplifications, which can compromise the quality of the MCA.

To address the limitations of filter-based MCAs (rigid structure, phase-lag) and MPC-based MCAs (high computational costs, necessary simplifications), neural network-based approaches have been proposed as these approaches could already show there advantages over traditional control algorithms in the different fields~\cite{2022Wurman, 2024Mandl}. These methods utilize machine learning techniques to model complex nonlinear dynamics with the fast execution time of an NN. As proposed in Figure~\ref{fig:taxonomy}, a separation between training the NN using supervised or reinforcement learning can be made. In the work of Konyuncu et al.~\cite{2020Koyuncu_CONF} a NN is trained to mimic the solution generated by an open-loop MPC algorithm, enabling faster, real-time execution by replacing the computationally expensive MPC with an NN approximation. This approach, which first optimizes motion with MPC and then trains the NN using SL, introduces an additional step. This step can be eliminated by directly training the NN to find the optimal solution using deep reinforcement learning (DRL), as this discipline bridges the gap between machine learning and control theory~\cite{2018Sutton_BOOK}.

Using an NN as a neural controller offers the advantage of avoiding the rigid structure inherent in filter-based MCAs, while also maintaining real-time capability, unlike NMPC-based MCAs, due to its computational efficiency during execution. Applying DRL to train the NN has demonstrated superior performance in generating control strategies for high-dimensional state spaces with complex objectives, particularly when a simulated model of the environment is available \cite{2020Bellemare_ARTC, 2022Degrave}. Moreover, DRL-trained controllers exhibit notably stable behavior even in the presence of significant deviations between the model and the real system~\cite{2021Lin_ARTC}.
Scheidel et al. firstly presented a proof of concept for an DRL based MCA, proposing a learning framework that operates in 2 degrees of freedom (DOF) in a linear workspace with limited training data~\cite{2024Scheidel_ARTC}. This framework was later extended to handle arbitrary trajectories while incorporating the human vestibular system in a subsequent study~\cite{2023Scheidel}. 
In the work presented here, the concept is further extended to develop a complete MCA that operates in six DOF within a nonlinear workspace, taking into account the kinematic and dynamic limitations of the MSP's joints. Thereby the first real time capable DRL based MCA with consideration of the robotic MSP's nonlinearities is introduced. The structure of the work is as follows: Section~\ref{sec:methodology} presents the methodology, which includes the DRL algorithm used, the proposed training framework, and the overall DRL based MCA. Section~\ref{sec:results} details the training procedure and provides a comprehensive validation against other MCA algorithms, as well as the application on the real system. The findings are summarized in the final conclusion and outlook in Section~\ref{sec:conclusion}.

\section{Methodology}~\label{sec:methodology}
To develop the DRL MCA, a state-of-the-art RL algorithm is used to train the NN. First, the theoretical background of RL and more specifically Proximal Policy Optimization, the utilized RL algorithm, is briefly outlined, followed by a description of the training framework. Finally, the outcome of the training process—the DRL based MCA—is presented.

\subsection{Theoretical Background of Reinforcement Learning}
The fundamental idea behind RL is to learn by interaction which is realized by the assumption of separating the world into an agent and an environment. The agent does not know the behavior of the  dynamic environment, but can interact with it via an defined interface and thereby learn its behavior. To achieve a mathematical definition of the sequential decision process the environment is described as a discrete-time stochastic Markov-decision-process (MDP). At time $t$ the environment is in a state $\boldsymbol{S}_t \in \mathcal{S}$, drawn from the set of states $\mathcal{S}$. Based on this state, the agent reacts with an action $\boldsymbol{A}_t \in \mathcal{A}$, from the set of actions $\mathcal{A}$, that leads to the next state $\boldsymbol{S}_{t+1}$. This consecutive flow leads to an trajectory $\tau=(\boldsymbol{S}_{0}, \boldsymbol{A}_{0}, \boldsymbol{S}_{1}, R_{1}, \boldsymbol{A}_{1}, \boldsymbol{S}_{2}, R_{2}, \boldsymbol{A}_{2},...)$ and the probability of this trajectory can be described as

\begin{equation}
	p_\theta (\tau) = \mu_0(\boldsymbol{S}_0) \prod_{t=1}^\infty \pi_\theta (\boldsymbol{A}_t | \boldsymbol{S}_t) \; T(\boldsymbol{S}_{t+1}|\boldsymbol{S}_t, \boldsymbol{A}_t).
\end{equation}

The transition probability $T(\boldsymbol{S}_{t+1}|\boldsymbol{S}_{t},\boldsymbol{A}_{t})$, which is unknown to the agent, describes the behavior of the environment by defining the probability of a stochastic transition for a given state $\boldsymbol{S}_{t}$ and action $\boldsymbol{A}_{t}$ to the next state $\boldsymbol{S}_{t+1}$. The policy $\pi_\theta (\boldsymbol{A}_t|\boldsymbol{S}_t)$ describes the behavior of the agent, defined by the probability of choosing the action $\boldsymbol{A}_{t}$ for a given state $\boldsymbol{S}_{t}$. For simple environments, the policy can be defined by look up table, mapping states to actions. For more complex cases, that occur in most real world applications, a function or a model like a NN are commonly used. Here, $\theta$ denotes the model parameters (weights and biases), which are adjusted during the training process. Lastly, $\mu_0(\boldsymbol{S}_0)$ denotes the start probability distribution of the environment.

Along with the state, the agent receives a numerical reward $R_t$, evaluating the value of the current state $\boldsymbol{S}_t$. The goal of the agent is to maximize the cumulative discounted reward $G(\tau)$ by the choice of the right trajectory $\tau$. The cumulative discounted reward is defined as

\begin{equation}
	G(\tau) = \sum_{t=0}^\infty \gamma^t R_{t+1},
\end{equation}

with the discount factor $0 \le \gamma \le 1$, defining the sensitivity of the agent to future rewards. During the training process the agent tries to adjust the parameters, $\theta$, of the policy, $\pi_\theta$, to maximize $G(\tau)$. Several approaches exist to achieve this~\cite{2018Sutton_BOOK}, but in this work proximal policy optimization (PPO), a policy gradient method by Schulman et al.~\cite{2017Schulman_ARTC}, is used and therefore further explained in the following. It was selected as it is a well-tested algorithm that has demonstrated competitive results across various robotic applications~\cite{2023Shakya_ARTC}. The basic concept of PPO is to alternate between sampling batches of training data based on the current policy, and optimizing the parameters of the differentiable policy in order to maximize $G(\tau)$. One challenge here is to find the right balance between exploration of potentially more rewarding, unknown territory of the environment and exploitation of the current policy with the risk of getting stuck in a local minima. The optimization of the policy is based on an estimator of the gradient of the expected reward with regard to the policy or respectively its parameters, $\theta$. Policy gradient methods estimated this gradient of the objective function $L(\theta)$ as

\begin{figure*}[t]
	\includegraphics[width=\textwidth]{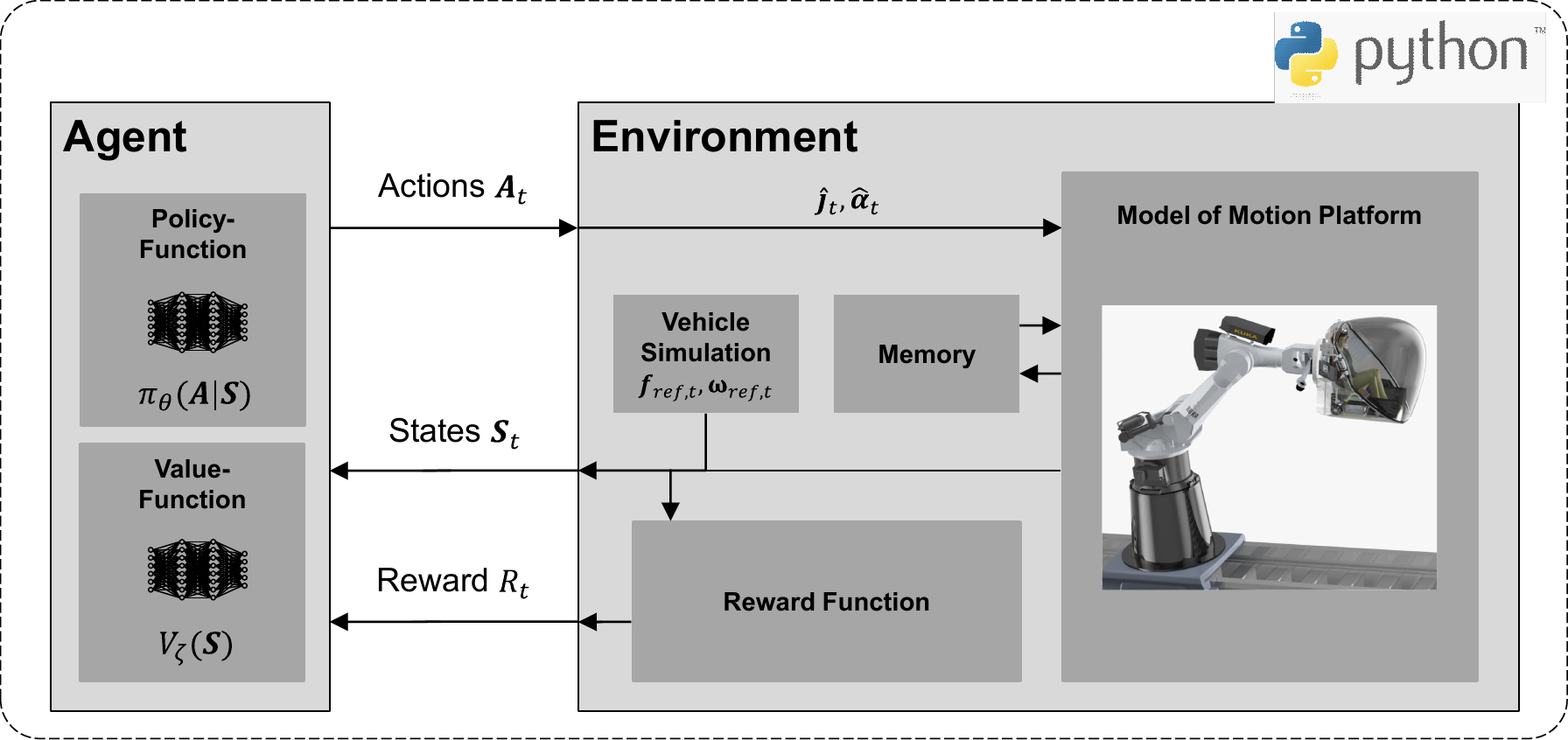}
	\caption{Schematic overview of the used training framework. It includes the nonlinear model of the motion platform and the reference trajectory coming from a vehicle simulation.} 
	\label{fig:framework}
\end{figure*}

\begin{equation}
	\nabla_\theta L (\theta) = \mathbb{E}_{\tau \sim p_\theta (\tau)}
	\Bigg[
	G(\tau) \sum_{t=0}^\infty \nabla_\theta \text{log} [\pi_\theta(\boldsymbol{A}_t|\boldsymbol{S}_t)]
	\Bigg].
\end{equation}

Here, $\mathbb{E}_{\tau \sim p_\theta (\tau)}$ denotes the expected value over a batch of trajectories drawn from $p_\theta(\tau)$. The reader is referred to fundamental works for further explanation of the necessary mathematical reformulation~\cite{2018Sutton_BOOK, 2020Dong}. The optimization of the parameters can then be realized by a stochastic gradient descent algorithm, as for example Adam~\cite{2014Kingma}. PPO, in an actor critic formulation, utilizes the advantage function $b(\boldsymbol{S}_t, \boldsymbol{A}_t)$ instead of $G(\tau)$. The advantage function expresses an estimation of the value of $\boldsymbol{S}_t$ and $\boldsymbol{A}_t$ over a baseline. Intuitively, it gives a measure of how good an action is in each specific state compared to the average value of that state. Its calculation is based on the value function

\begin{equation}
	V_\zeta(\boldsymbol{S}_t) = \mathbb{E}_\pi [G_t|\boldsymbol{S}_t],
\end{equation}

that is giving the expected discounted reward of a state for a policy $\pi$. The value function is stored in a second NN with the parameters $\zeta$, that is trained on the same batches of gathered training data as the policy function. Using the advantage function reduces the variance of the noisy training data and improves the stability and sample efficiency.

One challenge that policy gradient algorithms needs to solve is the balance between stability and efficiency of the policy optimization based on the current batch. PPO limits large policy updates by using a clipped objective function, reducing variance and preventing drastic changes to the policy.

\subsection{Proposed Framework for Training}~\label{subsec:training-framework}

The following section details the framework employed to achieve the successful training presented in this work. Figure~\ref{fig:framework} provides a schematic overview of the framework architecture. The interface between the agent and the environment is defined by the action vector $\boldsymbol{A}_t$ from the agent and the state vector $\boldsymbol{S}_t$ along with the reward $R_t$ from the environment. Due to the nature of the NN, it is crucial to normalize both vectors to ensure a stable training process. 

Similarly, as previously described by Scheidel et al.~\cite{2024Scheidel_ARTC}, the agent directly controls a kinematic model of the motion platform. Incorporating this model allows the agent to learn and adapt to the behavior and limitations inherent to the specific motion platform used. 
The second essential component of the framework is the vehicle simulation module. During the agent's training phase, this module provides the dynamics of a vehicle for pre-recorded trajectories as reference inputs, thus grounding the training process in realistic motion data. In the application phase of the motion cueing algorithm, the vehicle simulation functions as an interface to the vehicle's dynamic model, that generates reference trajectories in real-time. 
During training, the state of the motion simulator and the reference trajectory are passed from the environment to the agent as the state vector $\boldsymbol{S}_t$ and are also used as input to the reward function. The reward function evaluates the quality of the simulator motion mainly by comparing the motion of a simulator user with the reference motion of the vehicle user provided by the vehicle simulation. The reward $R_t$ is then also returned to the agent.

The following subsections provide a detailed explanation of each component, highlighting their roles, implementation details, and interactions within the framework.

\subsubsection{Motion Simulator} 
By including the full model, the agent learns to manage the kinematic properties and limitations of the platform, effectively adapting its control strategy to the specific characteristics of the simulator. The modular nature of this approach allows for the straightforward replacement of this component with a different simulator model if required. In this work, we utilize a model of the robotic motion platform located at the German Aerospace Center (DLR), as described by Bellmann et al. ~\cite{2011Bellmann_CONF}. The platform's nonlinear and non-cubic workspace, resulting from its serial kinematic arrangement, presents complex challenges to the control algorithm.

The control input provided by the agent consists of the suggested translational jerk $\hat{\boldsymbol{j}}_t^C$ in the $x$, $y$, and $z$ directions, and angular acceleration $\hat{\boldsymbol{\alpha}}_t^C$ in roll, pitch, and yaw of the end effector in the Cartesian coordinate system $C$ of the cockpit, forming an action vector $\boldsymbol{A}_t$ of length six. As demonstrated by Scheidel et al.~\cite{2024Scheidel_ARTC}, the use of the first derivatives of the dominantly perceived motion sensations - translational acceleration $\boldsymbol{a}_t$ and angular velocity $\boldsymbol{\omega}_t$ - significantly improves the training process. This can be explained with the exploration strategy employed by the reinforcement learning algorithm, which benefits from the smoothness and continuity in the states, and consequently, in the reward function provided by these derivative-based inputs.

The signal first needs to be rotated from the cockpit system~$C$ into the world system~$W$ using the inverse of the rotational matrix $\boldsymbol{T}_{t-1}^W$ of the last time step:

\begin{equation}
	\begin{bmatrix} 
		\hat{\boldsymbol{j}}_t^W \\ 
		\hat{\boldsymbol{\alpha}}_t^W 
	\end{bmatrix}
	=
	{\boldsymbol{T}_{t-1}^W}^{-1}
	\begin{bmatrix} 
		\hat{\boldsymbol{j}}_t^C \\ 
		\hat{\boldsymbol{\alpha}}_t^C 
	\end{bmatrix}.
\end{equation}

Next, $\hat{\boldsymbol{j}}_t^W$ needs to be integrated three times under consideration of the existing acceleration $\boldsymbol{a}_{t-1}^W$, velocity $\boldsymbol{v}_{t-1}^W$ and position $\boldsymbol{r}_{t-1}^W$ to receive the suggested new position of the end effector $\hat{\boldsymbol{r}}_t^W$ in the world frame. Similarly, $\hat{\boldsymbol{\alpha}}_t^W$ gets integrated using the existing angular velocity $\boldsymbol{\omega}_{t-1}^W$ to receive the suggested angular velocity $\hat{\boldsymbol{\omega}}_t^W$. Together with $\boldsymbol{T}_{t-1}^W$ this can be used to calculate the suggested new rotational matrix $\hat{\boldsymbol{T}}_{t}^W$. The integration are all implemented based on the Euler integration.

In order to reach the suggested orientation and position, the joint positions $\boldsymbol{q}_t$ are determined by solving the inverse kinematic problem of the motion platform. In this work, a 6-axis serial robot~\cite{2006Spong} is used and the six joint positions are defined as:

\begin{equation}
	\boldsymbol{q} = [q_1, q_2, q_3, q_4, q_5, q_6].
\end{equation}

Solving the inverse kinematic problem can be described reversely as finding the solution to

\begin{equation}
	\mathbb{F}(\boldsymbol{q}) = \boldsymbol{H},
\end{equation}

with $\boldsymbol{H}$ holding the given position and orientation of the robot's end-effector

\begin{equation}
	\boldsymbol{H} = 
	\begin{bmatrix} 
		\boldsymbol{T}^W & \boldsymbol{r}^W \\
		\boldsymbol{0} & 1 
	\end{bmatrix}.
\end{equation}

$\mathbb{F}$ describes the forward kinematic that can be analytical solved and always provides a unique solution for this type of robot. It consists of concatenation of matrix multiplication of the homogeneous transformation of the single joint

\begin{equation}
	\mathbb{F}(\boldsymbol{q}) = \prod_{i=0}^{6} G_i (q_i).
	\label{eq:forward_kin}
\end{equation}

The matrices $G_i(q_i)$, each corresponding to one joint, can be calculated using the Denavit-Hartenberg convention~\cite{2006Spong}.

For inverse kinematics of serial robots the process is more complex. Unlike forward kinematics, it is possible to have one solution, multiple solutions, or no solution at all. The primary objective in solving inverse kinematics is to find a joint configuration~$\boldsymbol{q}$ that achieves a position and orientation~$\boldsymbol{H}$ as close as possible to a desired target~$\hat{\boldsymbol{H}}$.

Given that changes in position and orientation between successive time steps are typically incremental, it is advantageous to search for a new joint configuration~$\boldsymbol{q}_t$ in the proximity of the previous configuration~$\boldsymbol{q}_{t-1}$. This continuity of motion allows for an analytical process that refines the current joint positions~$\boldsymbol{q}_t$ by solving for them in a way that minimizes the difference between~$\boldsymbol{H}_t$ and~$\hat{\boldsymbol{H}}_{t}$.

This means that applying inverse kinematics yields joint positions that closely match the commanded and achieved configurations $\boldsymbol{H}$, based on the robot's Denavit-Hartenberg parameters. However, it is still necessary to ensure that the MCA produces only physically feasible trajectories within the limits of the workspace and the kinematic constraints of the joints. To achieve this, limits must be imposed on the resulting joint acceleration $\boldsymbol{\ddot{q}}_t$, joint velocity $\boldsymbol{\dot{q}}_t$ and the joint position $\boldsymbol{q}_t$ for all six joints. These physical limits of the robot are implemented as:

\begin{equation}
	\ddot{q}_{\min,i} \leq \ddot{q}_{i,t} \leq \ddot{q}_{\max,i}, i \in 1...6
\end{equation}

\begin{equation}
	\dot{q}_{\min,i} \leq \dot{q}_{i,t} \leq \dot{q}_{\max,i}, i \in 1...6
\end{equation}

\begin{equation}
	q_{\min,i} \leq q_{i,t} \leq q_{\max,i}, i \in 1...6
\end{equation}

If it becomes necessary to limit the joint velocity, the resulting joint position must be recalculated by integrating the limited joint velocity. The same approach applies when limiting the joint acceleration. Additionally, a braking distance check is performed. This involves determining whether it is still possible to decelerate the joint to zero velocity before reaching the joint position limit, given the current joint position, current joint velocity, and joint acceleration constraints. If this is not feasible, the joint position must be adjusted accordingly.
After performing all checks, the result is a new, limited joint position $\boldsymbol{q}_{lim, t}$. The forward kinematics of the robot is then calculated using Equation~\ref{eq:forward_kin}, resulting in the physically feasible Cartesian position $\boldsymbol{r}^W_t$ and orientation $\boldsymbol{T}^W_t$ of the robot's end-effector.

Using values from the previous time step $t-1$, the Cartesian velocity $\boldsymbol{v}^W_t$, acceleration $\boldsymbol{a}^W_t$, and angular velocity $\boldsymbol{\omega}^W_t$, as well as the joint velocity $\boldsymbol{\dot{q}}_t$ and acceleration $\boldsymbol{\ddot{q}}_t$, are calculated through Euler differentiation. The rotation matrix is used to receive the Euler angles $\boldsymbol{\varphi}^W_t$, Cartesian velocity $\boldsymbol{v}^C_t$, acceleration $\boldsymbol{a}^C_t$, and angular velocity $\boldsymbol{\omega}^C_t$ in the coordinate system of the cockpit. Additionally, considering $\boldsymbol{T}^W_t$ and $\boldsymbol{a}_t$, the current specific force~$\boldsymbol{f}_t^C$ acting on the pilot is determined:

\begin{equation}
	\boldsymbol{f}^C_t = \boldsymbol{a}^C_t - \boldsymbol{T}^W_t \boldsymbol{g}^W.
\end{equation}

The specific force is particularly relevant for MCAs, as it combines the perceived effects of acceleration from translational motion with those of orientation changes relative to the gravitational vector, $\boldsymbol{g}$. Rather than directly replicating the translational acceleration of the reference vehicle, most MCAs aim to reproduce the specific force itself. Consequently, rather than attempting to replicate the translational acceleration of the reference vehicle, most MCAs focus on reproducing the specific force.
The motion platform model is implemented using the programming language C to ensure fast performance of the necessary calculations. The model is integrated with the framework using Cython~\cite{2011Behnel_ARTC}, with the kinematic state of the motion platform being passed for further processing and stored in a memory block for the next time step~$t+1$.

\subsubsection{Vehicle Simulator} 
Similar to the kinematic model of the motion platform, the vehicle simulation can also be viewed as a modular component that can be interchanged as needed for different specific use cases. The vehicle simulator used here is implemented in Modelica, an equation-based language for modeling of complex physical systems. The vehicle, a small passenger car, is controlled via an input devices such as a joystick, with commanded inputs including torque for acceleration, braking force for deceleration, and the steering angle. By modeling friction within the drive train, the maximum speed is limited by the equilibrium of forces. Additionally, a braking model is implemented to ensure wanted braking behavior.
To visualize the vehicle and the environment, the DLR Visualization 2 Library by Kümper et al.~\cite{2021Kuemper} is used. This library provides a real-time graphics environment for Modelica.

As described earlier, the physical values MCA's typically aims to reproduce are the specific force $\boldsymbol{f}_\text{ref}$ and the angular velocity $\boldsymbol{\omega}_\text{ref}$. To generate prerecorded input trajectories for the training process, these values are stored in log files, which can then be used as inputs to the environment. Before being further processed in the environment, the vectors are first pre-processed, including scaling and limiting, to ensure that the values remain within defined boundaries. 
During training, the provided trajectories define the operational domain within which the MCA is trained. In the application phase of the algorithm, the vehicle simulation transmits the reference values, $\boldsymbol{f}_{\text{ref},t}$ and $\boldsymbol{\omega}_{\text{ref},t}$ at each time step via a defined interface to the~MCA. 

\subsubsection{Reward Function}
The reference trajectory, characterized by the specific force and the angular velocity acting on the driver, and parts of the state of the motion platform are used as inputs to the reward function. The reward function consists of multiple components, and the selection and relative weighting of those are both challenging and crucial, as they define the optimization criteria and therefore the behavior of the agent. Shaping the reward function~\cite{1999Ng} is one of the common approaches in the process of designing the behavior of the agent, alongside modifying the MDP~\cite{2023Padalkar}. 

This task is particularly challenging in the context of an MCA because defining good behavior is inherently complex as different types of errors have varying effects on the simulator user.
As described by~\cite{1995Grant}, four distinct types of errors can be identified: false cues, missing cues, phase errors, and scale errors. Each of these errors impacts the perception of motion differently. Moreover, the perception of motion is influenced by subjective physiological and psychological factors. Due to the complexity of this topic, we refer to existing literature for a more detailed discussion~\cite{2021CasasYrurzum_ARTC, 1997Wu, 1995Grant}.

The reward function is designed to minimize these errors while ensuring a fast and stable training process. This is achieved by incorporating the following two components into the reward function: 1) minimize the error between the reference motion and the generated motion in specific force and angular velocity and 2) penalize the translational distance from the starting point to encourage washout of the motion. The reward function is defined as:

\begin{equation}
	\begin{split}
		R_t = 
		&- \sum_{i=1}^{3} \Bigg[ \lambda_f|f_{ref,i} - f^C_i|  + \lambda_{\omega}|\omega_{ref,i} - \omega^C_i|\Bigg] \\
		&- \sum_{i=1}^{3} \Bigg[ \lambda_r|r^W_i - r_{i,0}|\Bigg] \\
	\end{split}
\label{eq:reward-function}
\end{equation}

with $\boldsymbol{r}_0$ being the start position of the cockpit in the world system. and the operator $|\cdot|$ denoting the absolute value.

Since some components promote conflicting goals, selecting the appropriate weights, $\lambda_f, \lambda_\omega, \lambda_r$, is a challenging task. Here, the weights are determined through an optimization process, which is explained in more detail in section~\ref{subsec:training-setup}.

\subsection{Reinforcement Learning based MCA}

\begin{figure}
	\includegraphics[width=\columnwidth]{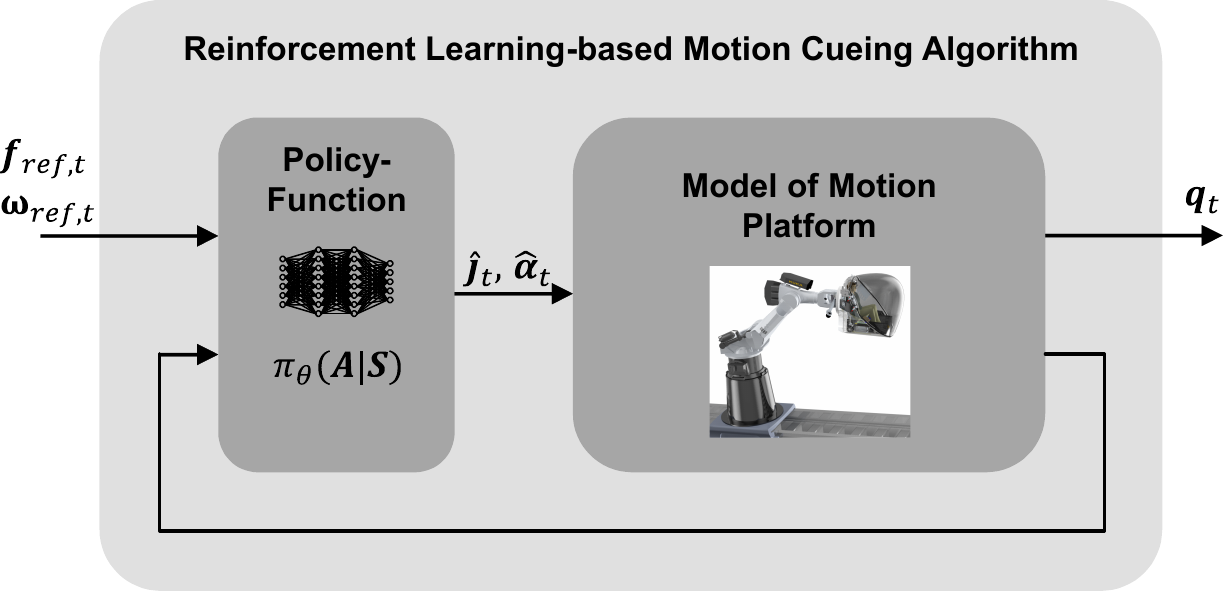}
	\caption{The processing steps of the RL-based MCA to generate a simulator joint trajectory from the vehicle dynamics simulation.} 
	\label{fig:RL-MCA}
\end{figure}

This subsection outlines the structure of the DRL-based MCA developed through the training process described in the previous subsection. As illustrated in Figure~\ref{fig:RL-MCA}, the MCA comprises two primary components: the policy function and the constraint model of the MSP. At each time step, the input to the DRL-MCA consists of $f_{ref,t}$ and $\omega_{ref,t}$, provided by the vehicle simulation, and the state of the MSP, provided by the mathematical model. These inputs form a vector that is processed by the trained policy function. The output of this NN is then passed to the motion platform’s constraint model, which is identical to the one used during the training process. Following the processing steps outlined in subsection~\ref{subsec:training-framework}, constraints on joint acceleration, velocity, and position ensure that only feasible joint trajectories, $q$, are provided to the simulator. The output is then applied to the MSP, which communicates the reached joint position back to algorithm, to close the control loop.

\section{Results and Discussion}~\label{sec:results}
In this section, the training setup, the validation and the application of the trained algorithm on the real hardware is discussed.

\subsection{Training Setup}~\label{subsec:training-setup}

\begin{table}[b]
	\centering
	\rowcolors{2}{gray!10!}{white}
	\caption{The adjusted hyper parameters used in the training. Names are according to the documentation of Stable Baselines3.}
	\label{tab:hyperparameter}
	\begin{tabular}{>{\bfseries} r r r}
		\textbf{Names} & \textbf{Original Value} & \textbf{Adjusted Value} \\
		\hline
		n steps & $2^{11}$ & $2^{17}$ \\
		batch size & $2^6$ & $2^{10}$ \\
		learning rate & $0.0003 $ & $0.000528$ \\
		gamma & $0.99$ & $0.996273$ \\
		n epochs & $10$ & $5$ \\
	\end{tabular}
\end{table}

The training is conducted using the RL library Stable Baselines3~\cite{2021Raffin_ARTC}, which provides open-source implementations of various RL algorithms, including PPO. The implementation utilizes the Gym library, which offers a standard API for RL applications~\cite{2016Brockman_ARTC}. The hyperparameters are selected based on the recommendations in the Stable Baselines3 documentation and are manually adjusted to fit the specific framework. The adjusted parameters are shown in Table~\ref{tab:hyperparameter}. Both the policy and value networks consist of two fully connected hidden layers with 1024 nodes per layer. The training is conducted over 100 million time steps, with a step size of 0.012 ms. With the presented hyperparameters one training takes approximately 24h running on one core of a Intel Xeon Gold 6136 without parallelization. 

The choice of reference trajectories presented to the agent during training has a significant impact on the training's success. Since the policy contains no inherent logic about the behavior of the MCA and is purely adjusted based on experience, the agent must be exposed to a sufficient number of extreme scenarios during training to ensure it performs well and remains stable across all possible input values. This necessitates that the reference trajectories be comprehensive enough to cover a wide variety of scenarios. Therefore, in addition to normal driving across all velocity ranges, the training data includes dynamic behaviors such as frequent harsh braking, maximum acceleration, constant driving in circles, slalom driving and no also no driving. The total length of the training data comprises approximately 84 minutes. 
To ensure that the agent can generalize its learned behavior to inputs beyond the training data, a similarly complex testing trajectory is recorded. This testing data has a total length of 5 minutes. To evaluate the performance of the agent on this test trajectory quantitative criteria for the objective assessment of the MCA needs to be defined. Two main metrics are used in this work: the root-mean-squared  error (RMSE) defined as

\begin{equation}
	\epsilon(\boldsymbol{k}_{ref}, \boldsymbol{k}) = \sqrt{\frac{1}{T} \sum_{t=1}^{T}(k_{t,ref}-k_t)^2}
\end{equation}

and Pearson's correlation coefficient (PCC) defined as

\begin{equation}
	\rho(\boldsymbol{k}_{ref}, \boldsymbol{k}) = \frac{\mathbb{E}[(k_t-\mu_k)(k_{t,ref}-\mu_{k_{ref}})]}{\sigma_k \sigma_{k_{ref}}}.
\end{equation}

Here, $T$ is the overall length of the trajectory, $\mathbb{E}$ the expected value, $\mu$ the mean and $\sigma$ the standard deviation. The PCC was first used in the context of MCA's by Asadi et al.~\cite{2016Asadi_ARTC} and gives a measure of the similarity of the shape of two signals. Generally, $\boldsymbol{k}$ and $\boldsymbol{k}_{ref}$ represent two vectors that need to be compared. In the present work, both, $\epsilon$ and $\rho$, are determined for the reference values, $\boldsymbol{f}_{\text{ref},t}$ and~$\boldsymbol{\omega}_{\text{ref},t}$, versus the values acting on the simulator user, $\boldsymbol{f}_t^C$ and~$\boldsymbol{\omega}^C_t$. If a complete similarity of two signals is reached the RMSE goes to $0$ and the PCC to $1$. Therefore, an overall objective function of 

\begin{equation}\label{eq:ojective_function}
	\begin{split}
		\min \sum_{i=1}^{6} \Bigg[ \epsilon(\boldsymbol{k}_{ref,i}, \boldsymbol{k}_i) - \rho(\boldsymbol{k}_{ref,i}, \boldsymbol{k}_i) \Bigg] \\
		\text{with } \boldsymbol{k} = \{\boldsymbol{f}_x, \boldsymbol{f}_y, \boldsymbol{f}_z, \boldsymbol{\omega}_x, \boldsymbol{\omega}_y, \boldsymbol{\omega}_z\}
	\end{split}
\end{equation}

can be formulated, with a minimal value $-6$. In order to achieve equal consideration of the RMSE of the angular velocity and the specific force, which have different units, the values must be of a similar order of magnitude. This is achieved if the specific force is considered in $m/s^2$ with a weighting of $1$ and the angular velocity in $deg/s$ with a weighting of $0.1$. Based on the evaluation on this objective function of the performance of the agent for the test trajectory, the agent for further validation is chosen. Additionally, the objective function is used to optimize the weights ($\lambda_f, \lambda_\omega, \lambda_r$) of the reward function, as given in Equation~\ref{eq:reward-function}. This is accomplished using the optimization framework Optuna~\cite{2019Akiba}, which provides implementations of various optimization algorithms for automated, parallelized hyperparameter search. Due to sample efficiency, Tree-structured Parzen Estimator (TPE) is chosen as optimization algorithm~\cite{2011Bergstra}. Performing $500$ runs on a parallelized system results in a set of weights for the reward function that leads to optimized value of the objective function, as given in Equation~\ref{eq:ojective_function}.

\begin{table*}
	\setlength{\tabcolsep}{4pt}
	\setlength\extrarowheight{5pt}
	\centering
	\rowcolors{5}{gray!10!}{white}
	\caption{The objective value for all four algorithms applied to 15 validation trajectories. The objective value is composed of the RMSE and the PCC of the specific force and the angular velocity in all 3 dimensions and needs to be minimized. The best possible value is -6 for a perfect match between reference and generated trajectory.}
	\label{tab:metrics}
	\begin{tabular}{>{\bfseries} r | r r r r r r r r r r r r r r r }
		\hline
		\multirow{2}{*}{\textbf{Algorithms}} & \multicolumn{15}{c}{\textbf{Trajectories}} \\ \cline{2-16}
		& \textbf{1} & \textbf{2} & \textbf{3} & \textbf{4} & \textbf{5} & \textbf{6} & \textbf{7} & \textbf{8} & \textbf{9} & \textbf{10} & \textbf{11} & \textbf{12} & \textbf{13} & \textbf{14} & \textbf{15}\\
		\hline
		CW MCA & -0.050 & 2.379 & 1.482 & 1.382 & 1.652 & -1.191 & 0.830 & 0.702 & 0.301 & 1.786 & 2.910 & 3.201 & 2.551 & 1.181 & 2.875 \\
		NMPC MCA & -1.284 & \textbf{0.598} & -0.326 & \textbf{-0.339} & -0.676 & \textbf{-2.369} & \textbf{-1.037} & -0.660 & -0.743 & 0.329 & \textbf{1.296} & \textbf{1.318} & \textbf{0.608} & \textbf{-1.046} & 0.561 \\
		DRL MCA & \textbf{-1.460} & 0.605 & \textbf{-0.346} & 0.036 & \textbf{-1.026} & -2.197 & -0.518 & \textbf{-0.950} & \textbf{-1.167} & \textbf{0.083} & 1.618 & 1.940 & 0.714 & -0.578 & \textbf{0.423} \\
	\end{tabular}
\end{table*}

\subsection{Algorithms for Validation}
In this chapter, the state-of-the-art algorithms, including CW  and nonlinear MPC (NMPC) that are being used for validation are presented.

The CW MCA is based on the work of Reid and Nahon~\cite{1985Reid_ARTC}. The inputs to the algorithm are the scaled and limited specific force and angular velocity. Both signals are filtered using high-pass filters to isolate the high-frequency components, which are directly applied to the motion platform. Additionally, low-pass filters are used to extract the low-frequency components of the specific force, which are used for tilt coordination. The CW algorithm operates without considering the current state or dynamic limitations of the motion platform. Avoiding these limitations relies solely on limiting and scaling the input signals and tuning the filter parameters. To optimize performance, a global optimization process based on simplicial homology global optimization (SHGO)~\cite{2022Scheidel} is employed. The objective function for optimization is equivalent to Equation~\ref{eq:ojective_function}, and the same training trajectory data is utilized. While this tuning procedure results in a better performing parameter set, it does not guarantee that the algorithm will respect the dynamic constraints of the platform. The algorithm is implemented in Python.

The NMPC based MCA used here is based on the work by Katliar et al.~\cite{2018Katliar}. The nonlinear plant consists of two sets of equations: the state transition equations and the output equations. The state transition model consists of a double integrator per joint. The input of each integrator is the joint acceleration, and the states are the joint position and velocity. The output equations consist of the forward kinematics of the robot, its first two derivatives, and tilt coordination. The plant outputs are the angular velocity and tilt-coordinated linear acceleration at the location of the head of the simulator driver. With this formulation the underlying optimization problem is designed to find the joint acceleration trajectories that minimize inertial reference tracking error and joint states along the prediction horizon subject to all joint position, velocity, and acceleration limits. The main advantage of this method is that it is guaranteed to produce feasible joint trajectories. However, the computational cost of solving a nonlinear program at every control iteration poses a significant challenge for use in real-time. The implementation of this algorithm was done in Python using HILO-MPC~\cite{ 2024Pohlodek_ARTC} and CasADi~\cite{2018Andersson_ARTC}. A prediction and control horizon length of 200 steps is used for the plant step size of 12~ms.

\subsection{Results of Validation}

To assess the performance of the DRL MCA, a comprehensive comparison with the CW and NMPC MCAs is conducted. For this purpose, 15 different validation files, each spanning 51 seconds, are recorded. These files include a range of driving scenarios, such as normal driving, harsh braking and acceleration, slalom driving, and continuous circular driving. The 15 validation files serve as input for all four algorithms. To ensure a realistic comparison, the trajectories generated by each MCA are applied to a model of the robotic MSP. This model limits the joint positions, velocities, and accelerations to reproduce a realistic behavior of the MSP and is similar to the model used during the training of the DRL MCA and in the optimization of the NMPC MCA. In case of strong violations of the limitations in joint positions, velocities or acceleration this post processing reduces the quality of the motion sensation severely as a strong deviation in suggested and physically possible trajectory is induced. Therefore, CW in combination with the necessary post processing, achieves the best possible result with a conservative tuning, as only minor changes to the suggested trajectory become necessary.

The specific force and angular velocity experienced by the simulator user are compared against the reference values. A qualitative assessment is performed by calculating the RMSE and PCC for each of the three dimensions of specific force and angular velocity. This results in a total of 12 metrics per file and per algorithm, which are then aggregated into a single objective function, as defined in Equation~\ref{eq:ojective_function}. The significance of an evaluation of the algorithm via the objective function is limited, as a lot of information is lost when summarizing the trajectory to one scalar value. For example, the result depends on the chosen weights and a nuanced consideration of the errors is fundamentally difficult in quantitative terms~\cite{2024Kolff}. Nevertheless, a basic categorization of the quality of the algorithms and an initial assessment of the potential of DRL based MCA is possible.

All objective values are presented in Table~\ref{tab:metrics}. To gain a more comprehensive understanding of the results, a Wilcoxon Signed-Rank test was performed. This statistical test is a non-parametric alternative to the paired t-test and is used to compare two paired samples. The test was conducted to compare the performance of the DRL MCA against both other algorithms, allowing for an assessment of the DRL MCA’s performance. For both comparisons, the null hypothesis states that there is no significant difference of the objective values between the two algorithms being compared. The significance level ($\alpha$) for rejecting the null hypothesis was set to $0.05$. The comparison DRL MCA to CW MCA resulted in a p-value of $6.104\cdot10^{-5}$, indicating clearly that the null hypothesis can be rejected, and a significant difference in performance exists between DRL MCA and the washout based algorithm. As shown in Table~\ref{tab:metrics}, the DRL MCA achieved better objective values than CW MCA across all 15 trajectories. The third Wilcoxon Signed-Rank test compared the DRL MCA to the NMPC MCA and resulted in a p-value of $0.561$. This indicates that the null hypothesis cannot be rejected for this test, and no significant difference in performance was found between these two algorithms. The three conducted Wilcoxon Signed-Rank tests lead to the conclusion that CW MCA and LMPC MCA show significant worse result than DRL MCA. NMPC shows similar performances based on the analyzed metric.

\begin{figure}
	\includegraphics[width=\columnwidth]{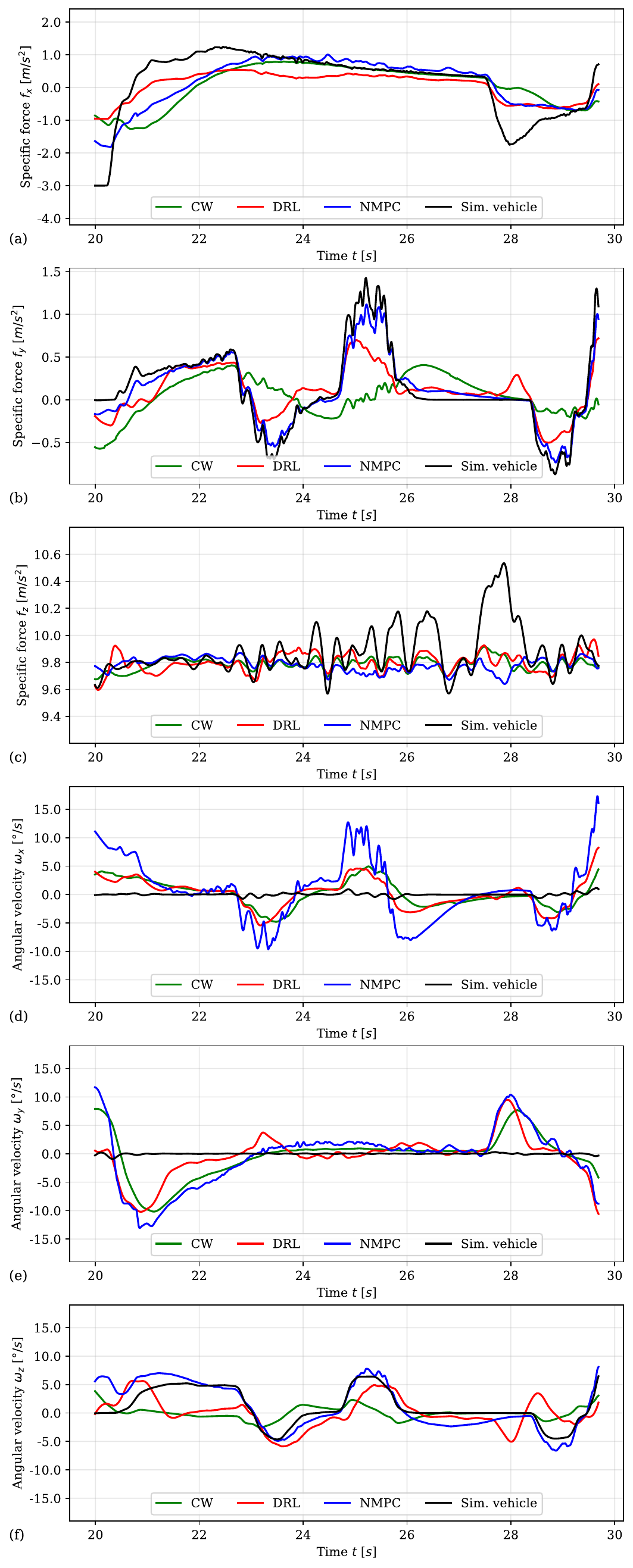}
	\caption{Specific force and angular velocity for the validation trajectory 3. Presented are the values of the reference and the output of the NMPC, CW and DRL MCAs.} 
	\label{fig:f_and_w_plot}
\end{figure}

Next, a more detailed assessment of one single trajectory is conducted. Trajectory 3 is chosen here as it provides comparably similar objective values for NMPC and DRL MCA. Figure~\ref{fig:f_and_w_plot} provides a qualitative assessment by comparing the specific forces and angular velocities produced by the four algorithms to the reference trajectory for validation trajectory~3. Combined with Table~\ref{tab:rms-errors} and Table~\ref{tab:ccs}, presenting the RMSEs and PCCs for this trajectory across all four algorithms, a more detailed evaluation is possible. 
Comparing the performance of DRL MCA and NMPC MCA on the angular velocities, $\boldsymbol{\omega}_x$ and $\boldsymbol{\omega}_y$, against the performance on $\boldsymbol{f}_x$ and $\boldsymbol{f}_y$ it becomes clear that the amount of tilt coordination differs strongly. This impression gets confirmed by looking at the RMSE and the PCC: 
NMPC MCA achieves a better match in $\boldsymbol{f}_x$ and $\boldsymbol{f}_y$ by performing more tilt coordination by rotational movement of the simulator cabin, which leads to stronger deviations in $\boldsymbol{\omega}_x$ and $\boldsymbol{\omega}_y$. Conversely, DLR MCA performs less tilt coordination, so the angular velocities show a better match but the specific forces are followed less closely. This is a typical trade off in the design of MCAs and especially for driving simulations, that come with a more dynamic acceleration profile than flight simulation, the preferred rate of tilt coordination is highly subjective and an optimal value is not determined~\cite{2009Fischer, 2024Kolff}. 
The performance of CW on the trajectory is unsatisfactory as it is only able to follow the reference with deviation in the phase, particularly noticeable for $\boldsymbol{f}_x$ and $\boldsymbol{f}_y$. 
This can be assigned to a conservative tuning that becomes necessary as the path planning is confined to the Cartesian workspace without consideration of the limitations of the MSP's joints but is also an inherent issue of the filter based approach.

\begin{figure}
	\includegraphics[width=\columnwidth]{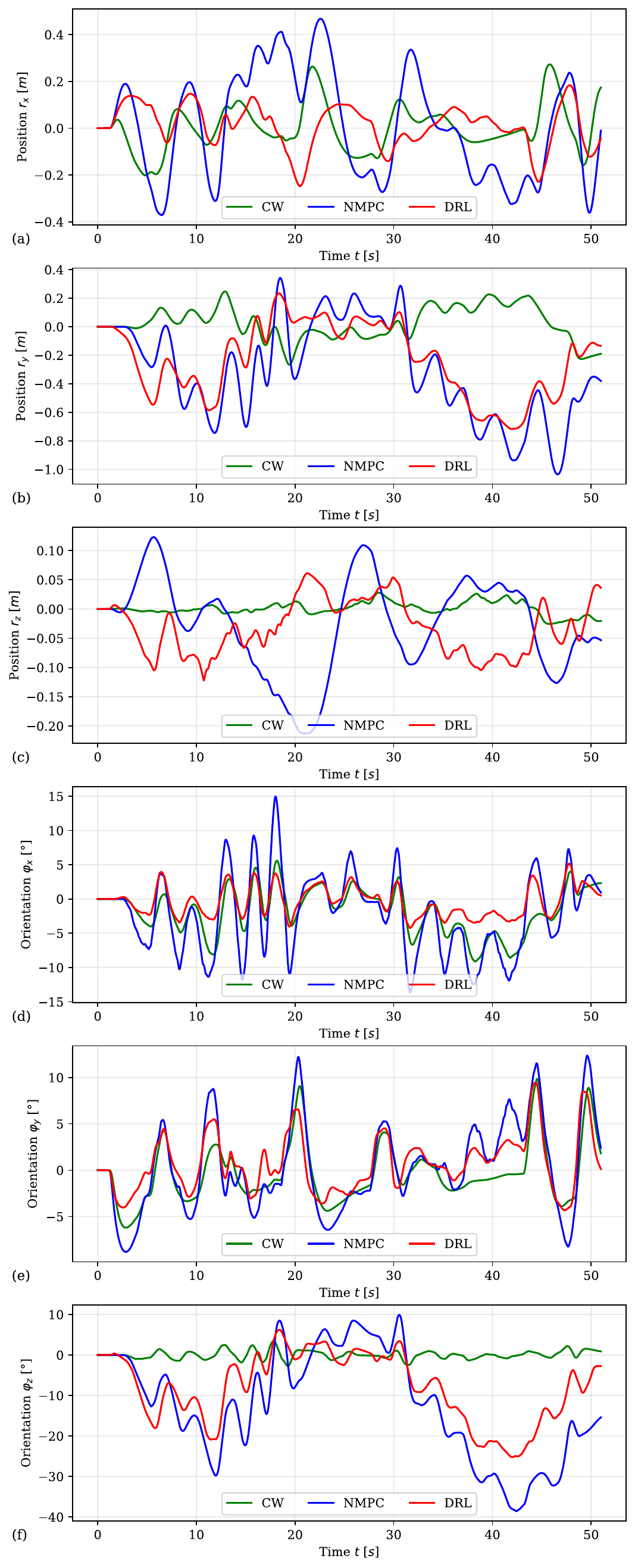}
	\caption{Position and orientation of the end effector for the validation trajectory~3. Presented are the values  of NMPC, CW and DRL MCAs.} 
	\label{fig:r_and_phi_plot}
\end{figure}

Figure~\ref{fig:r_and_phi_plot} illustrates the position and orientation of the MSP's end effector for trajectory 3 across all three algorithms. It is evident that NMPC utilizes tilt coordination the most, as indicated by the significant changes in orientation. Overall, NMPC makes greater use of the workspace compared to the other two algorithms. This highlights that the DRL MCA achieves comparable results for specific force and angular velocity while utilizing the resources of the MSP more efficiently.

\begin{table}
	\centering
	\setlength\extrarowheight{2pt}
	\rowcolors{5}{gray!10!}{white}
	\caption{Comparison of the rms-errors of the specific forces and the angular velocities for trajectory 3. The best possible value is~$0.0$ for identical trajectories.}
	\begin{tabular}{>{\bfseries} c|c c c|c c c}
		\hline
		\multirow{2}{*}{\textbf{Algorithms}} & \multicolumn{3}{c|}{\textbf{RMSE [$\text{ms}^{-2}$]}} & \multicolumn{3}{c}{\textbf{RMSE [$\text{s}^{-1}$]}} \\ \cline{2-7}
		& $\boldsymbol{f}_x$ & $\boldsymbol{f}_y$ & $\boldsymbol{f}_z$ & $\boldsymbol{\omega}_x$  & $\boldsymbol{\omega}_y$ & $\boldsymbol{\omega}_z$\\ \hline
		CW MCA & 0.879 & 1.000 & \textbf{0.178} & 0.423 & \textbf{0.382} & 0.606 \\ \hline
		NMPC MCA & \textbf{0.630} & \textbf{0.233} & 0.224 & 0.972 & 0.547 & \textbf{0.411} \\ \hline
		DRL MCA & 0.667 & 0.872 & 0.221 & \textbf{0.347} & 0.410 & 0.472 \\ \hline
	\end{tabular}
	\label{tab:rms-errors}
\end{table}

\begin{table}
	\centering
	\setlength\extrarowheight{2pt}
	\rowcolors{5}{gray!10!}{white}
	\caption{Comparison of the correlation coefficient of the specific forces and the angular velocities for trajectory 3. The best possible value is $1.0$ for a perfect correlation.}
	\begin{tabular}{>{\bfseries}c|c c c|c c c}
		\hline
		\multirow{2}{*}{\textbf{Algorithms}} & \multicolumn{6}{c}{\textbf{Correlation Coefficient [-]}} \\ \cline{2-7}
		& $\boldsymbol{f}_x$ & $\boldsymbol{f}_y$ & $\boldsymbol{f}_z$ & $\boldsymbol{\omega}_x$  & $\boldsymbol{\omega}_y$ & $\boldsymbol{\omega}_z$\\ \hline
		CW MCA & 0.508 & 0.437 & \textbf{0.489} & 0.093 & 0.070 & 0.386 \\ \hline
		NMPC MCA & 0.806 & \textbf{0.989} & -0.066 & \textbf{0.429} & \textbf{0.292} & \textbf{0.895} \\ \hline
		DRL MCA & \textbf{0.907} & 0.779 & 0.247 & 0.377 & 0.284 & 0.741 \\ \hline
	\end{tabular}
	\label{tab:ccs}
\end{table}

\begin{table}[b]
	\centering
	\rowcolors{0}{gray!10!}{white}
	\caption{Comparison of the average computational time needed per time step. The control frequency is $12$ ms, marking the threshold for real-time capability.}
	\begin{tabular}{>{\bfseries}c|c c c}
		\hline
		Algorithms & CW & NMPC & DRL \\ \hline
		Average Time [ms] & \textcolor{DarkGreen}{0.300} & \textcolor{red}{1546.5} & \textcolor{DarkGreen}{0.965} \\ \hline
	\end{tabular}
	\label{tab:computational_time}
\end{table}

Looking at the computational load, as presented in Table~\ref{tab:computational_time}, the strength of the DRL MCA becomes clear. The average computational time per time step of $12$ ms is $0.965$ ms ($\sigma = 0.501$ ms), making the algorithm clearly real time capable. NMPC MCA, being the only other algorithm which considers the model of the MSP and therefore is generating a feasible trajectory for the robotic simulator, reaches an average value of $1546.5$ ms ($\sigma = 460.7$ ms) per $12$ ms time step. This high computational load makes it not real time capable and not applicable to the real system. With $0.300$ ms CW is also able to meet the real-time requirements, due to its comparatively simple structure, which does not take into account the nonlinear dynamics of the MSP. However, as shown in the previous discussion, it is not able to achieve compatible performance in this setup.

From the conducted validation, it can be concluded that DRL is capable of delivering competitive motion signals compared to state-of-the-art algorithms, while being the only one among the four evaluated algorithms that simultaneously considers the kinematic constraints of the robotic MSP and meets the real-time requirements of the system.

\subsection{Sim-To-Real: Implementation on real Motion Platform}

To complete the evaluation, the DRL MCA is applied on the real system to perform initial safety and feasibility checks for future studies. The system which is used is the DLR Robotic Motion Simulator of the German Aerospace Center~\cite{2011Bellmann_CONF}. As shown in Figure~\ref{fig:dlr_rms}, it consists of a KUKA KR500/2 TÜV industrial robot with 6 DOF attached to a 10 meter linear axis for additional workspace. For the test, only the 6 DOF of the robotic kinematic, and not the linear axis, are utilized. The algorithm as shown in Figure~\ref{fig:RL-MCA} is implemented using a Python environment. Due to the low computational load it is possible to run the simulator in a closed loop setup. At every time step the simulator receives a joint position, that can be reached without violating any physical constraints. From the point of view of DRL, and machine learning in general, this sim-to-real-transfer from a simulated model of the robot to a real system comes with additional challenges~\cite{2020Zhao}, due to a reality gap caused by the modeling of the system.

The safety and feasibility checks are performed with a driver in the loop, steering the vehicle from inside cabin of the motion simulator. After a driving period of 20 minutes the algorithm behaved as intended and no indicators of motion sickness could be observed. This simple test shows that a transfer from the simulation to the real robotic system is possible. However, in order to be able to carry out a more precise assessment of the MCA, a comprehensive study with standardized questionnaires, sufficient test subjects and comparison algorithms would have to be carried out.
\section{Conclusion}~\label{sec:conclusion}
We have introduced the first nonlinear DRL based MCA designed to control a MSP  and enhance the quality of motion simulation. Existing algorithms either are not able to consider the kinematic and dynamics properties of the MSP, struggling therefore with workspace nonlinearities, (e.g. CW MCA, LMPC MCA) or lack real-time capability due to high computational demand (e.g NMPC MCA). The here presented DRL based MCA overcomes these limitations by incorporating nonlinear kinematics and dynamics while achieving real-time capability, facilitated by the separation of computationally intensive offline training and efficient online application. 
The training framework involves an agent controlling a simulated robotic MSP, learning through trial and error the effects of control inputs on system states. To ensure comprehensive state-space coverage, the agent is exposed during training to a wide variety of input signals, representing the full range of states in the coupled vehicle simulation. Training alternates between environment interaction, based on the agent’s evolving knowledge, and policy and value function improvement using PPO on the recorded data. Optimization of the policy and value function, both implemented as neural networks, is driven by a reward function designed to enhance motion perception. 
A validation study conducted on 15 validation files compared the RL MCA with CW, LMPC, and NMPC algorithms, with the RL MCA showing competitive performance across all files and achieving the highest performance on seven of them. Finally, successful application of the DRL MCA to the DLR Robotic Motion Simulator demonstrates its capability to control the simulator in real time with human pilots actively operating the MSP from within the cockpit. 

Further studies are needed for a comprehensive understanding of the DRL MCA’s performance. A comparative analysis with other real-time MCAs on the physical motion simulator, involving multiple participants and standardized methodologies, is essential for a full assessment. Future work could pursue several directions, including enhancing the environment by integrating a mathematical model of the human vestibular system or incorporating predictions of pilot behavior, which could increase the algorithm's capabilities. Evaluating DRL MCA performance on diverse tracks and vehicle types and exploring advancements in the fields of RL and artificial intelligence, such as new training algorithms or new model architectures, may also lead to improvements in both training quality and efficiency.

\bibliography{../../01_General/00_Bib_Files/Paper}
\end{document}